# Thermodynamics and Relativity: A Message to Physics Teachers
## (for an easier understanding of the thermodynamic theory)

Jean-Louis Tane (tanejl@aol.com). November 2007
Formerly with the Department of Geology, University Joseph Fourier, Grenoble, France

**Abstract**: Whoever has to learn or to teach thermodynamics is confronted with conceptual difficulties which are specific to this field of physics ([1],[2]). It seems that they can be eliminated by inserting relativity in the thermodynamic theory. The aim of this paper is to summarize the problem and draw attention upon a fundamental point of the discussion.

**- 1 - Summarized explanation of the problem.**

Let us consider a thermodynamic system defined as a given amount of gas contained in a vessel. We suppose that the gas can exchange work (dW) with the surroundings thanks to a mobile piston, of negligible weight, situated at the upper part of the vessel. It can also exchange heat (dQ) with a thermostat situated at the lower part, from which it is separated by a diathermic wall.

If the gas evolves from an initial state $(P_1, V_1, T_1)$ to a final state $(P_2, V_2, T_2)$, three remarks can be made about the thermodynamic analysis of the process.

### 1.1 First remark

Concerning the work exchange, the basic equation we have to use is:

$$dW = -P_e dV \qquad (1)$$

which corresponds to the general case of an irreversible process. For a reversible process, it takes the form:

$$dW = -P_i dV \qquad (2)$$

In these equations, $P_e$ means "external pressure" and $P_i$ "internal pressure".

A preliminary conclusion is that, for a given value of dV, the difference between $dW_{irr}$ and $dW_{rev}$ is expressed by the general equation:

$$dW_{irr} - dW_{rev} = dV(P_i - P_e) \qquad (3)$$

Observing that dV (which means $dV_{gas}$) is positive when $P_i > P_e$ and negative when $P_i < P_e$, it can be noted that the term $dV(P_i - P_e)$ is always positive, so that we have in all cases:

$$dW_{irr} > dW_{rev} \qquad (4)$$

Another writing of eq. 4 is therefore:

$$dW_{irr} = dW_{rev} + dW_{add} \qquad (5)$$

where $dW_{add}$ means $dW_{additional}$ and has a positive value.

### 1.2 Second remark

In the usual conception of thermodynamics, the change in internal energy of the system, designated as dU, is supposed to have the same value, whatever the level of irreversibility of the process. This proposition constitutes the first law of thermodynamics and implies the equality:

$$dU_{irr} = dU_{rev} \tag{6}$$

Since dU is itself given by the relation:

$$dU = dQ + dW \tag{7}$$

we see that, depending on whether the process is reversible or irreversible, eq. 7 takes the respective forms:

$$dU_{rev} = dQ_{rev} + dW_{rev} \tag{8}$$

$$dU_{irr} = dQ_{irr} + dW_{irr} \tag{9}$$

To conciliate the condition $dU_{irr} = dU_{rev}$ (eq. 6) with the condition $dW_{irr} > dW_{rev}$ (eq. 4), the only possible solution seems to be given by the proposition:

$$dQ_{irr} < dQ_{rev} \tag{10}$$

### 1.3 Third remark

When a condensed phase (solid or liquid) is heated, it is implicitly admitted that the amount of heat has the same value whatever the level of irreversibility of the heating process. Indeed, the amount of heat (ΔQ) is calculated, in all cases, by integration of the classical relation:

$$dQ = mcdT \tag{11}$$

where m, c and dT represent the mass, the specific heat capacity and the change in temperature of the phase, respectively.

When the phase taken in consideration is a gas, a distinction is made between $dQ_p$ and $dQ_v$, and explicit equations exist permitting the calculation of quantities such as $Q_p - Q_v$.

Concerning the difference between $dQ_{irr}$ and $dQ_{rev}$, the discussion remains open, as explained below, but an important information can be extracted from the equation expressing the second law of thermodynamics. This equation is generally written as:

$$dS = dQ/T + dS_i \tag{12}$$

whose precise meaning is :

$$dS = dQ/T_e + dS_i \qquad (13)$$

where $dS_i$ has a positive value, except in the limited case of a reversible process for which it becomes zero.

Eq. 13 has the dimension of an entropy, but takes the dimension of an energy if it is written under the form:

$$T_e dS = dQ + T_e dS_i \qquad (14)$$

Knowing that $dS_i$ is positive and that $T_e$ is positive too (being an absolute temperature), we see that the product $T_e dS_i$ is necessarily positive. At first glance, we can be tempted to think that, in eq.14, $dQ$ has the significance $dQ_{irr}$ and $T_e dS$ the significance $dQ_{rev}$, so that this equation appears to be in perfect agreement with eq.10 ($dQ_{irr} < dQ_{rev}$).

Although this interpretation is the one implicitly admitted in conventional thermodynamics, it is not totally convincing because $dQ_{rev}$ is defined as $dQ_{rev} = T_i dS$, not as $dQ_{rev} = T_e dS$. Knowing that $dS = dS_e + dS_i$ and that, consequently, $T_e dS = T_e dS_e + T_e dS_i$, it can be observed that, in eq.14, the first meaning of the term $dQ$ is $dQ = T_e dS_e$. Having noted just above that $dQ = T_i dS$, we are led to the conclusion that $T_e dS_e = T_i dS$, so that eq. 14 can also be written:

$$T_e dS = T_i dS + T_e dS_i \qquad (15)$$

whose meaning is therefore:

$$dQ_{irr} = dQ_{rev} + T_e dS_i \qquad (16)$$

As a consequence eq. 16 can be given the form:

$$dQ_{irr} > dQ_{rev} \qquad (17)$$

or the form:

$$dQ_{irr} = dQ_{rev} + dQ_{add} \qquad (18)$$

where $dQ_{add}$ has a positive value.

Obviously, the result inferred from equations 16 and 17 is not the one expected from eq.10 ($dQ_{irr} < dQ_{rev}$), but the opposite. This is the sign that we are confronted to a real problem which needs to be solved.

### - 2 - Suggested solution of the problem

The conceptual difficulty just encountered comes from an incompatibility between eq.10 ($dQ_{irr} < dQ_{rev}$) and eq. 14 ($T_e dS = dQ + T_e dS_i$). Since eq. 10 is itself directly derived from eq. 6 ($dU_{irr} = dU_{rev}$), a possible explanation is that the postulate expressed by eq. 6 is an inadequate interpretation of the first law of thermodynamics and needs to be substituted by the postulate:

$$dU_{irr} \neq dU_{rev} \qquad (19)$$

We have seen with eq. 5 (work exchange) and eq. 18 (heat exchange) that the difference between an irreversible and a reversible process takes the respective forms:

$$dW_{irr} = dW_{rev} + dW_{add} \qquad (5)$$

$$dQ_{irr} = dQ_{rev} + dQ_{add} \qquad (18)$$

where both $dW_{add}$ and $dQ_{add}$ are positive.

Remembering that $dU = dQ + dW$ (eq.7), the suggestion which comes in mind is that $dU$ obeys itself a general equation which can be written:

$$dU_{irr} = dU_{rev} + dU_{add} \qquad (20)$$

where $dU_{add}$ is positive. This is equivalent to say that, in all cases, the relation between $dU_{rev}$ and $dU_{irr}$ is:

$$dU_{irr} > dU_{rev} \qquad (21)$$

Compared with the classical conception of thermodynamics, eq. 20 covers both the extended interpretation of the first law ($dU_{irr} \neq dU_{rev}$) and the energetic transcription of the second law ($dU_{add} > 0$) which is itself an extension of its usual entropic writing ($dS_i > 0$).

The question raised by eq. 20 is evidently the origin of the energy corresponding to the term $dU_{add}$. An hypothesis advanced in previous papers ([3],[4]) suggests that it can be explained be a disintegration of mass occurring within the system, according to the Einstein mass-energy relation $E = mc^2$. In this conception, $dU_{add}$ is considered as having the significance:

$$dU_{add} = dE = -c^2 dm \qquad (22)$$

the sign minus corresponding to the fact that a decrease in mass results in an increase in energy and conversely.

Due to this reference to the mass-energy relation eq. 20 can be written:

$$dU_{irr} = dU_{rev} - c^2 dm \qquad (23)$$

and appears as a general connection between thermodynamics and relativity.

## 3. Conclusions

As indicated in its title, this paper is mainly addressed to physics teachers and the salient feature of the discussion is the one presented in section 1.3. It explains that there is a real incompatibility between eq. 10 ($dQ_{irr} < dQ_{rev}$) and eq. 14 ($T_e dS = dQ + T_e dS_i$) due to the fact that in this last equation, $dQ_{rev}$ is represented by $dQ$ and $dQ_{irr}$ by $T_e dS$, not the reverse. Although this interpretation has not been evoked, until now, in books of thermodynamics, it seems that many authors are not fully satisfied with the usual theory. The references about this matter listed below ([1], [2]) are limited to books of thermodynamics intended for geologists, but the problem evoked is general and inherent to physics, not to geology.

The solution suggested above strengthens the idea that inserting relativity (Einstein's relation) in the reasoning makes thermodynamics more easily understandable.

Referring to the classical conception of thermodynamics, it can be noted that the positive value of $dS_i$ (internal increase in entropy) is not at all contested. It is interpreted as the symptom of a positive value of $dU_i$ (internal increase in energy) designated, in the present paper, under the symbol $dU_{add}$ (additional energy, whose value is positive)

As a consequence, the natural evolution of a system, classically understood as an increase in entropy, becomes an increase in energy, itself linked to a decrease in mass. It can be emphasized that this proposition concerns systems exclusively made of inert matter, such as those on which the thermodynamic theory was stated in the XIXth century. It is not excluded that it needs to be inverted for systems made – at least partially – of living matter ([5],[6]).

From the practical point of view, the well-known efficiency of the thermodynamic tool is not contested either. Referring to an elementary exchange of energy, a numerical example has been examined [3], showing the difference between the classical interpretation and the new suggested one.

**Acknowledgements:** I would like to thank the readers who contacted me about my previous papers. Several of them being teachers, my hope is that the matter discussed in the present article could be of some help for their courses.